\documentclass[aps,prl,superscriptaddress,amssymb,reprintm,twocolumn,nobibnotes]{revtex4-1}
\pdfoutput=1
\usepackage{amsmath, amsfonts}    
\usepackage{graphicx}   
\usepackage{verbatim}   
\usepackage{color}      
\usepackage{subfigure}  
\usepackage{hyperref}   
\usepackage{natbib}

\graphicspath{{./figures/}}

\newcommand{\ket}[1]{\left|#1\right\rangle}

\newcommand{\up}{\uparrow}
\newcommand{\down}{\downarrow}

\newcommand{\ccite}[1]{\textsuperscript{#1}}
\newcommand{\micro}{$\mathrm{\mu}$}
\newcommand{\suppl}{\mbox{Supplementary} \mbox{Information}}

\begin{document}

\title{Experimental loophole-free violation of a Bell inequality using entangled electron spins separated by 1.3 km}
\author{B. Hensen}
\author{H. Bernien}
\altaffiliation{Present address: Department of Physics, Harvard University, Cambridge, Massachusetts 02138, USA}
\author{A.E. Dr\'{e}au}
\author{A. Reiserer}
\author{N. Kalb}
\author{M.S. Blok}
\author{J. Ruitenberg}
\author{R.F.L. Vermeulen}
\author{R.N. Schouten}
\affiliation{QuTech, Delft University of Technology, P.O. Box 5046, 2600 GA Delft, The Netherlands}
\affiliation{Kavli Institute of Nanoscience Delft, Delft University of Technology, P.O. Box 5046, 2600 GA Delft, The Netherlands}

\author{C. Abell{\'a}n}
\author{W. Amaya}
\author{V. Pruneri}
\affiliation{ICFO-Institut de Ciencies Fotoniques, Av. Carl Friedrich Gauss, 3, 08860 Castelldefels, Barcelona, Spain.}
\author{M.W. Mitchell}
\affiliation{ICFO-Institut de Ciencies Fotoniques, Av. Carl Friedrich Gauss, 3, 08860 Castelldefels, Barcelona, Spain.}
\affiliation{ICREA-Instituci{\'o} Catalana de Recerca i Estudis Avançats, Lluis Companys 23, 08010 Barcelona, Spain}

\author{M. Markham}
\author{D.J. Twitchen}
\affiliation{Element Six Innovation, Fermi Avenue, Harwell Oxford, Didcot, Oxfordshire OX110QR, United Kingdom.}

\author{D. Elkouss}
\author{S. Wehner}
\affiliation{QuTech, Delft University of Technology, P.O. Box 5046, 2600 GA Delft, The Netherlands}
\author{T.H.\ Taminiau}
\author{R. Hanson}
\email{r.hanson@tudelft.nl}
\affiliation{QuTech, Delft University of Technology, P.O. Box 5046, 2600 GA Delft, The Netherlands}
\affiliation{Kavli Institute of Nanoscience Delft, Delft University of Technology, P.O. Box 5046, 2600 GA Delft, The Netherlands}

\begin{abstract}
For more than 80 years, the counterintuitive predictions of quantum theory have stimulated debate about the nature of reality\ccite{1}. In his seminal work\ccite{2}, John Bell proved that no theory of nature that obeys locality and realism can reproduce all the predictions of quantum theory. Bell showed that in any local realist theory the correlations between distant measurements satisfy an inequality and, moreover, that this inequality can be violated according to quantum theory. This provided a recipe for experimental tests of the fundamental principles underlying the laws of nature. In the past decades, numerous ingenious Bell inequality tests have been reported\ccite{3-12}. However, because of experimental limitations, all experiments to date required additional assumptions to obtain a contradiction with local realism, resulting in loopholes\ccite{12-15}. Here we report on a Bell experiment that is free of any such additional assumption and thus directly tests the principles underlying Bell's inequality. We employ an event-ready scheme\ccite{2,16,17} that enables the generation of high-fidelity entanglement between distant electron spins. Efficient spin readout avoids the fair sampling assumption (detection loophole\ccite{13,14}), while the use of fast random basis selection and readout combined with a spatial separation of 1.3 km ensure the required locality conditions\ccite{12}. We perform 245 trials testing the CHSH-Bell inequality\ccite{18} $S \leq 2$ and find $S = 2.42 \pm 0.20$. A null hypothesis test yields a probability of $p = 0.039$ that a local-realist model for space-like separated sites produces data with a violation at least as large as observed, even when allowing for memory\ccite{15,19} in the devices. This result rules out large classes of local realist theories, and paves the way for implementing device-independent quantum-secure communication\ccite{20} and randomness certification\ccite{21,22}.
\end{abstract}

\maketitle

We consider a Bell test in the form proposed by Clauser, Horne, Shimony and Holt (CHSH)\ccite{18} (Fig. 1a). The test involves two boxes labelled A and B. Each box can accept a binary input ($0$ or $1$) and subsequently delivers a binary output ($+1$ or $-1$). In each trial of the Bell test a random input bit is generated on each side and input to the respective box. The random input bit triggers the box to produce an output value that is recorded. The test concerns correlations between the output values (labeled $x$ and $y$ for box A and B, respectively) and the input bits (labeled $a$ and $b$ for A and B respectively) generated within the same trial.

The remarkable discovery made by Bell is that in any theory of physics that is both local (physical influences do not propagate faster than light) and realistic (physical properties are defined prior to and independent of observation) these correlations are bounded more strongly than in quantum theory. In particular, if the input bits can be considered free random variables (condition of ``free will'') and the boxes are sufficiently separated such that locality prevents communication between the boxes during a trial, the following inequality holds under local realism:
\begin{align}\label{eq:S_paramter}
S &= \left< x \cdot y \right>_{\left(0,0\right)} + \left< x \cdot y \right>_{\left(0,1\right)} \\\nonumber
  & \hspace{1cm} + \left< x \cdot y \right>_{\left(1,0\right)} - \left< x \cdot y \right>_{\left(1,1\right)} \leq 2
\end{align}
where $\left< x \cdot y \right>_{\left(a,b\right)}$  denotes the expectation value of the product of $x$ and $y$ for input bits $a$ and $b$. 

Quantum theory predicts that the Bell inequality can be significantly violated in the following setting. We add one particle, for example an electron, to each box. The spin degree of freedom of the electron forms a two-level system with eigenstates $\ket{\up}$ and $\ket{\down}$. For each trial, the two spins are prepared into the entangled state $\ket{\psi^-}=\left(\ket{\up\down}-\ket{\down\up}\right)/\sqrt{2}$. The spin in box A is then measured along direction $\mathrm{Z}$ (for input bit $a = 0$) or $\mathrm{X}$ (for $a = 1$) and the spin in box B is measured along $(-\mathrm{Z}+\mathrm{X})/\sqrt{2}$ (for $b = 0$) or $(-\mathrm{Z}-\mathrm{X})/\sqrt{2}$ (for $b = 1$). If the measurement outcomes are used as outputs of the boxes, quantum theory predicts a value of $S=2\sqrt{2}$, showing that the combination of locality and realism is fundamentally incompatible with the predictions of quantum mechanics.

\begin{figure*}
	\includegraphics[width=163mm]{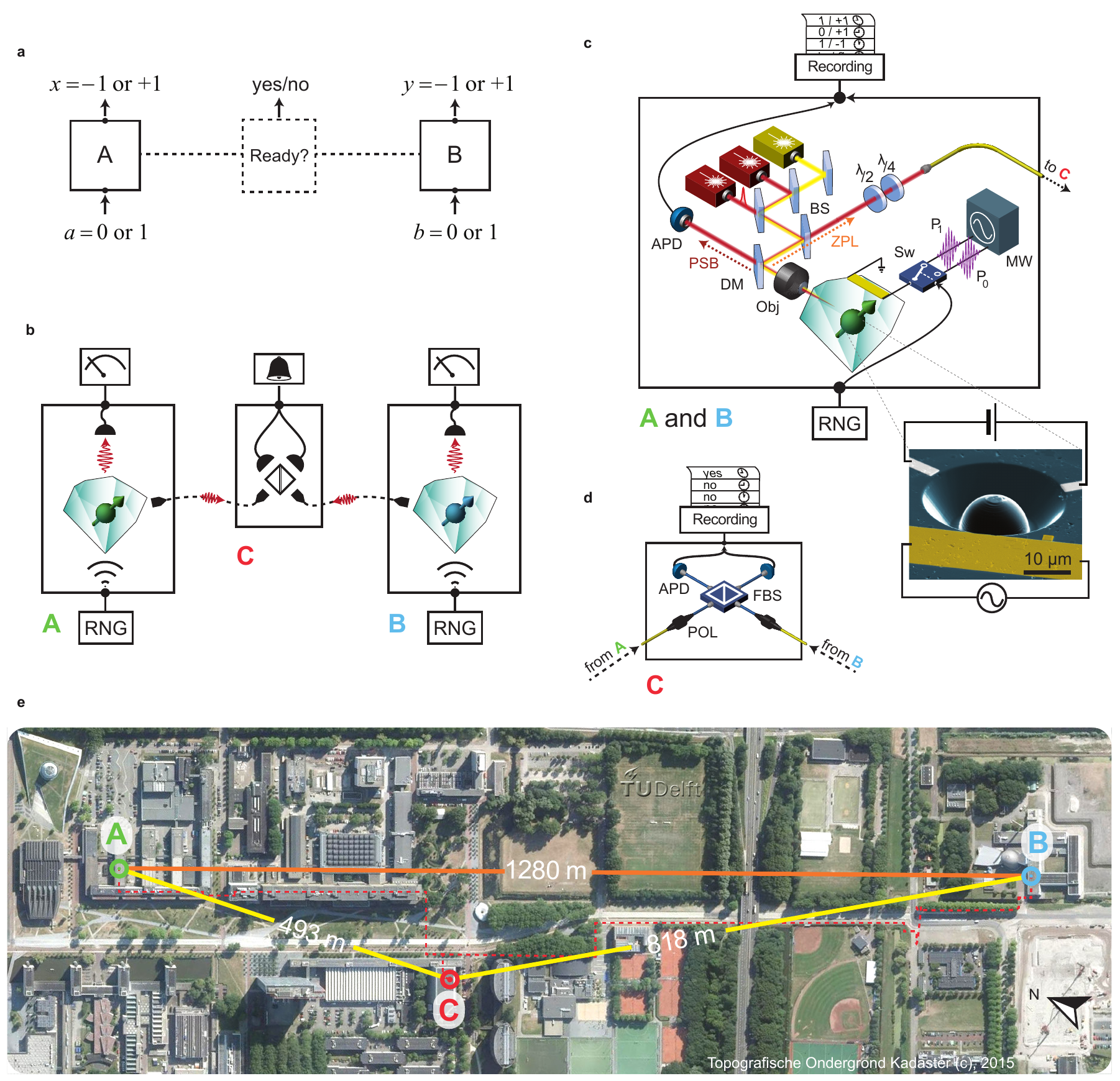}
	\caption{\label{Bell_fig1} \textbf{Bell test schematic and experimental realisation.} \textbf{(a)} Bell test setup: Two boxes, A and B, accept binary inputs $(a,b)$ and produce binary outputs $(x,y)$. In an event-ready scenario, an additional box C gives a binary output signalling that the boxes A and B were successfully prepared. \textbf{(b)} Experimental realisation. The setup comprises three separate laboratories, A, B and C. The boxes at location A and B each contain a single NV centre electron spin in diamond. A quantum random number generator (RNG) is used to provide the input to the box. The spin is  read out in a basis that depends on the input bit and the resulting signal provides the output of the box.  A third box at location C records the arrival of single photons that were previously emitted by, and are entangled with, the spins at A and B. The detection of two such photons constitutes the event-ready signal. \textbf{(c)} Detailed experimental setup at A and B. The electronic spin associated with a single nitrogen vacancy (NV) centre in diamond is located in a low temperature confocal microscope setup (Obj). A fast switch (Sw) transmits only one out of two different microwave (MW) pulses ($\mathrm{P_0}$ and $\mathrm{P_1}$), depending on the output of a quantum random number generator (RNG). The microwave pulses are then applied via a gold strip-line deposited on the diamond surface (inset, scanning electron microscope image of a similar device). The optical frequencies of the NV are tuned by a d.c. electric field applied to on-chip gate electrodes (inset). Pulsed red and yellow lasers are used to resonantly excite the optical transitions of the NV centre. The emission (dashed arrows) is spectrally separated into an off-resonant part (phonon side band, PSB) and a resonant part (zero-phonon line, ZPL), using a dichroic mirror (DM). The PSB emission is detected with a single-photon counter (APD), whose detection events are recorded together with the generated random numbers by a time-tagging device. The ZPL emission is mostly transmitted through a beam-sampler (BS, reflection $\leq 4$\%) and two wave plates ($\lambda/2$ and $\lambda/4$), after which it is coupled to a single mode fibre that guides the light to location C. \textbf{(d)} Setup at location C. The single-mode fibres from locations A and B are connected to the input ports of a fibre-based beam splitter (FBS) after passing a fibre-based polarizer (POL). Photons in the two output ports are detected using single photon counters, and detection events are recorded. \textbf{(e)} Aerial photograph of the campus of Delft University of Technology indicating the distances between locations A, B and C. The red dotted line marks the path of the fibre-connection.}
\end{figure*}	

 Bell's inequality provides a powerful recipe for probing fundamental properties of nature: all local realistic theories that specify where and when the free random input bits and the output values are generated can be experimentally tested against it.
 
Violating Bell's inequality with entangled particles poses two main challenges: excluding any possible communication between the boxes (locality loophole\ccite{12}) and guaranteeing efficient measurements (detection loophole\ccite{13,14}). First, if communication is possible, a box can in principle respond using knowledge of \emph{both} input settings, rendering the Bell inequality invalid. The locality conditions thus require the boxes A and B and their respective free input bit generations to be separated in such a way that signals travelling at the speed of light (the maximum allowed under Special Relativity) cannot communicate the local input setting of box A to box B, before the output value of box B has been recorded, and vice versa. Also, the input bits should not be able to influence the preparation of the entangled state. Second, disregarding trials in which a box does not produce an output bit (i.e. assuming fair sampling) would allow the boxes to select trials based on the input setting. The fair sampling assumption thus opens a detection loophole\ccite{13,14}: the selected subset may show a violation even though the set of all trials may not. 

The locality loophole has been addressed with pairs of photons separated over a large distance\ccite{4} in combination with fast random number generators to provide the input settings\ccite{5,9}. However, these experiments left open the detection loophole due to imperfect detectors and inevitable photon loss during the spatial distribution of entanglement. The detection loophole has been closed in different experiments\ccite{6-8,10,11}, but these did not close the locality loophole. To date, no experiment has closed all the loopholes simultaneously.

Because of unclosed loopholes, Bell's inequality could not be tested in previous experiments without introducing additional assumptions. Therefore, a Bell test that closes all experimental loopholes at the same time  - commonly referred to as a loophole-free Bell test\ccite{14,17,23} is of foundational importance to the understanding of nature. In addition, a loophole-free Bell test is a critical component for device-independent quantum security protocols\ccite{20} and randomness certification\ccite{21,22}. In such adversarial scenarios all loopholes must be closed, since they allow for security breaches in the system\ccite{24}.

An elegant approach for realizing a loophole-free setup was proposed by Bell himself\ccite{2}. The key idea is to record an additional signal (dashed box in Fig. 1a) to indicate whether the required entangled state was successfully shared between A and B, i.e. whether the boxes were ready to be used for a trial of the Bell test. By conditioning the validity of a Bell test trial on this event-ready signal, failed entanglement distribution events are upfront excluded from being used in the Bell test. 
	
We implement an event-ready Bell setup\ccite{16,17} with boxes that employ the electron spin associated with a single NV defect centre in a diamond chip (Fig 1b). The diamond chips are mounted in closed-cycle cryostats ($T = 4$K) located in distant laboratories named A and B (Fig. 1c). We control the spin state of each electron with microwave pulses applied to on-chip striplines ({Fig. 1c,~inset}). The spins are initialized through optical pumping and read out along the $Z$ axis via spin-dependent fluorescence\ccite{25}. The readout relies on resonant excitation of a spin-selective cycling transition (12 ns lifetime), causing the NV centre to emit many photons when it is in the bright $m_s=0$ spin state, while it remains dark when it is in either of the $m_s=\pm 1$ states. We assign the output value +1 ($m_s=0$) in case we record at least one photo-detector count during the readout window, and the output value -1 ($m_s=\pm 1$) otherwise. Readout in a rotated basis is achieved by first rotating the spin followed by readout along $\mathrm{Z}$.

We generate entanglement between the two distant spins by entanglement swapping\ccite{16} in the Barrett-Kok scheme\ccite{26,27} using a third location C (roughly midway between A and B, see Fig. 1e). First we entangle each spin with the emission time of a single photon (time-bin encoding). The two photons are then sent to location C, where they are overlapped on a beam-splitter and subsequently detected. If the photons are indistinguishable in all degrees of freedom, the observation of one early and one late photon in different output ports projects the spins at A and B into the maximally entangled state $\ket{\psi^-}=\left(\ket{\up\down}-\ket{\down\up}\right)/\sqrt{2}$, where $m_s=0\equiv\ket{\up}$, $m_s=-1\equiv\ket{\down}$. These detections herald the successful preparation and play the role of the event-ready signal in Bell's proposed setup. As can be seen in the space-time diagram in Fig.~2a, we ensure that this event-ready signal is space-like separated from the random input bit generation at locations A and B.

\begin{figure*}
	\includegraphics[width=148mm]{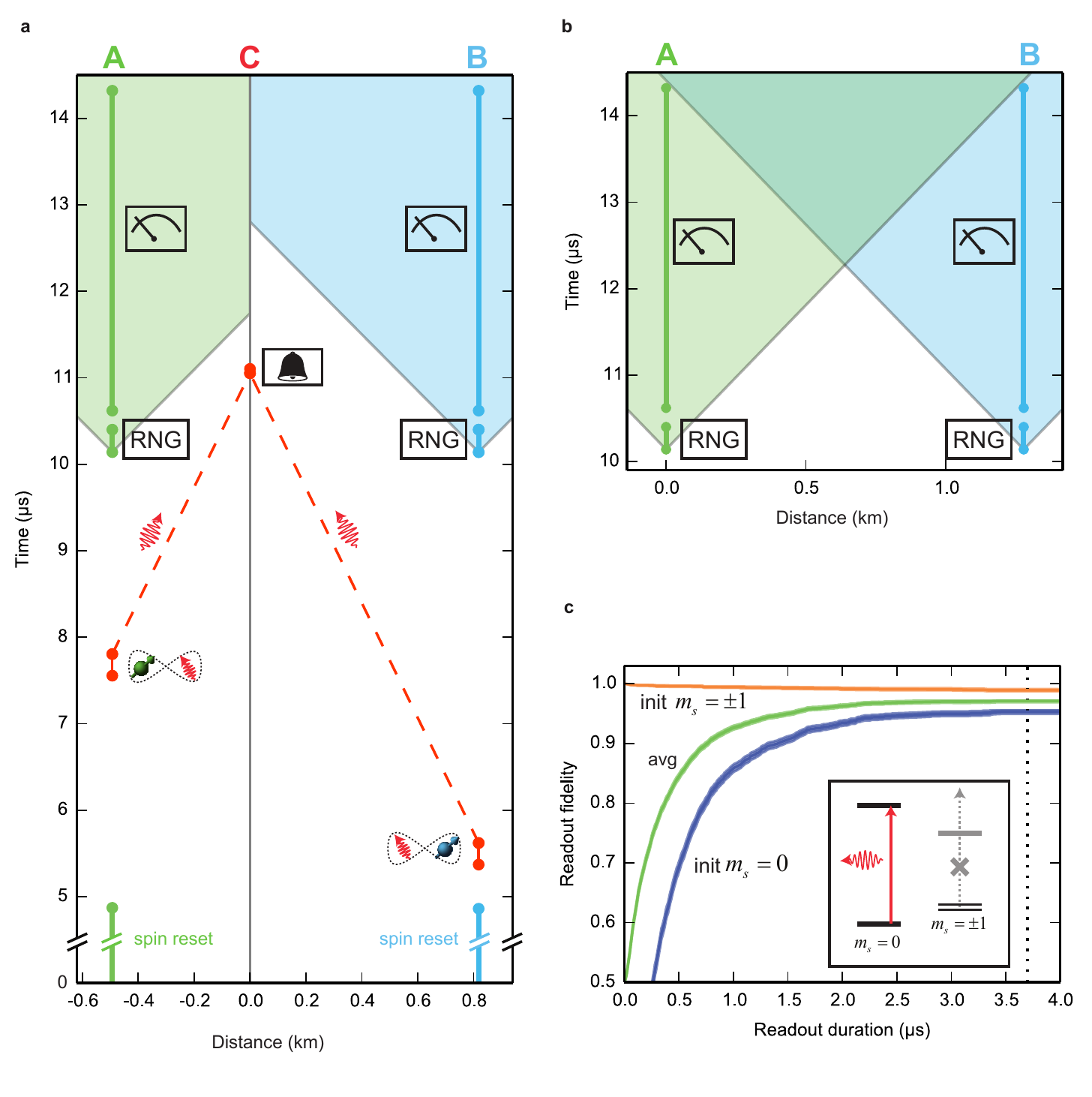}
	\caption{\label{Bell_fig2} \textbf{Space-time analysis of the experiment.} \textbf{(a)} Space-time diagram of a single repetition of the entanglement generation. The x-axis denotes the distance along the lines A-C and C-B. First, the spins at location A and B are initialised by optical pumping. Then spin-photon entanglement is generated, at A and at B, timed such that the two photons from A and B arrive simultaneously at location C where the detection time of the photons is recorded.  Successful preparation of the spins is signalled (bell symbol) by a specific coincidence detection pattern. Independent of the event-ready signal, the setups at location A and B choose a random basis (RNG symbol), rotate the spin accordingly and start the optical spin-readout (measurement symbol). Vertical bars indicate the duration of the spin reset, random number generation and measurement operations. The event ready signal lies outside the future light-cone (coloured regions) of the random basis choice of A and B (see main text for details). \textbf{(b)} Space-time diagram of the Bell test. The x-axis denotes the distance along the line A-B. The readout on each side is completed before any signal traveling at the speed of light can communicate the basis choice from the other side. The uncertainty in the depicted event times and locations is much smaller than the symbol size. \textbf{(c)} Single-shot spin readout fidelity at location A as a function of readout duration (set by the latest time a detection event recorded in the time-tagger is taken into account). Blue (orange) line: fidelity of outcome +1 (-1) when the spin is prepared in $m_s = 0$. ($m_s = \pm 1$). Green line: average readout fidelity. Dotted line: Readout duration used (3.7 \micro s). The inset shows the relevant ground and excited state levels (not to scale).}
\end{figure*}

The separation of the spins by 1280 m defines a 4.27~\micro s time window during which the local events at A and B are space-like separated from each other (see the space-time diagram depicted in Fig. 2b). To comply with the locality conditions of the Bell test, the choice of measurement bases and the measurement of the spins should be performed within this time window. For the basis choice we employ fast random number generators with real-time randomness extraction\ccite{28}. We reserve 160 ns for the random basis choice, during which time one extremely random bit is generated from 32 partially random raw bits (\suppl). The random bit sets the state of a fast microwave switch that selects one out of two pre-programmed microwave pulses implementing the two possible readout bases (Fig. 1c). Adding the durations of each of the above steps yields a maximum time from the start of the basis choice to the start of the readout of 480 ns. We choose the readout duration to be 3.7~\micro s, leaving 90 ns to cover any uncertainty in the distance between the laboratories and the synchronization of the setup (estimated total error is at most 16 ns, see \suppl). For this readout duration, the combined initialisation and single-shot readout fidelity of sample A is $(97.1 \pm 0.2)$\% (Fig 2c); sample B achieves $(96.3 \pm 0.3)$\%. In summary, the use of the event-ready scheme enables us to exploit highly entangled spin states and simultaneously comply with the strict locality conditions of the Bell setup, while the single-shot nature of the spin readout closes the detection loophole.

\begin{figure}[t]
	\includegraphics[width=83mm]{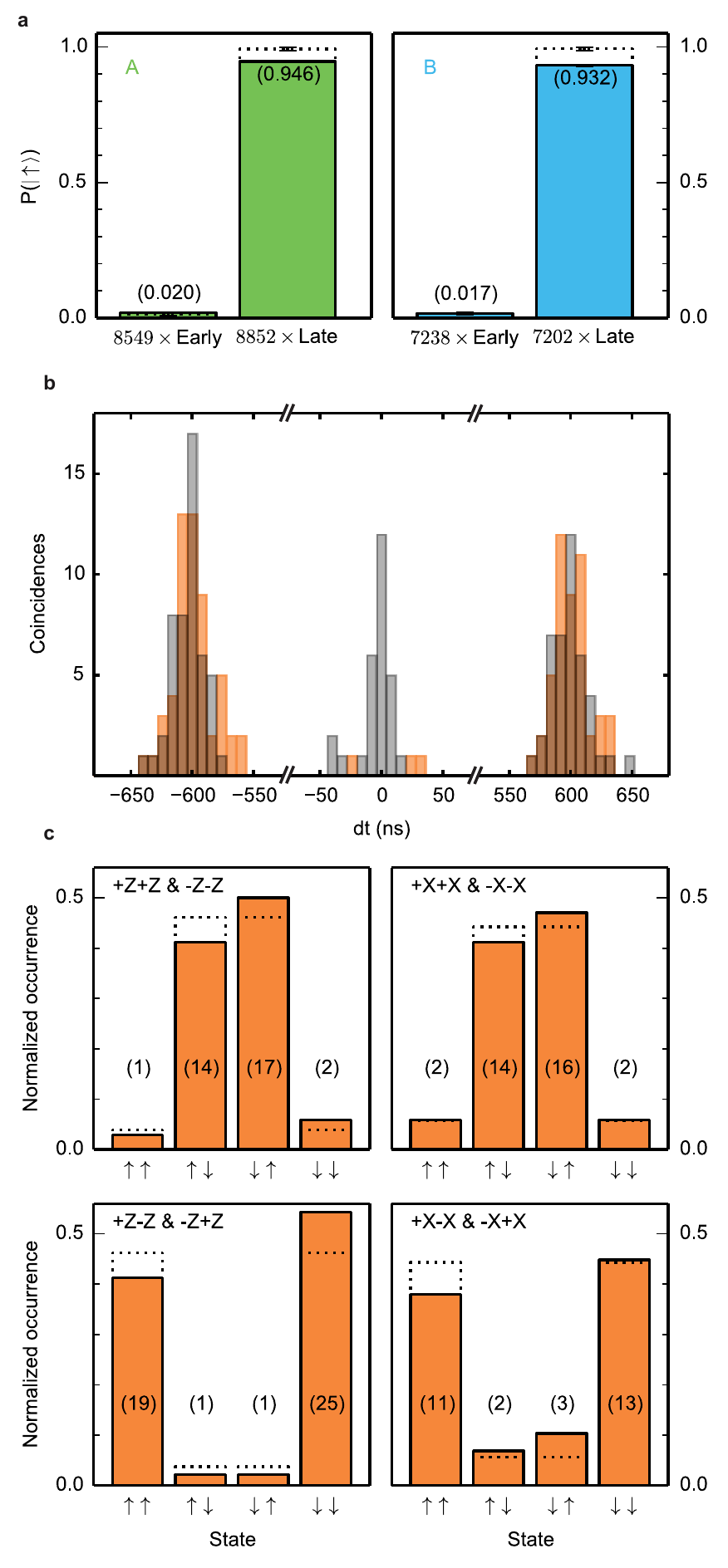}
\end{figure}

Before running the Bell test we first characterize the setup and the preparation of the spin-spin entangled state.  Figure 3a displays correlation measurements on the entangled spin-photon states to be used for the entanglement swapping. For both location A and B we observe near-unity correlations between spin state and photon time bin when spin readout errors are accounted for. We then estimate the degree of indistinguishability of the single photons emitted at locations A and B in a two-photon interference\ccite{29}  experiment at location C, i.e. after the photons have travelled through a combined total of 1.7 km of single-mode optical fiber. Using the observed two-photon interference contrast of $0.90 \pm 0.06$ and the spin-photon correlation data we estimate that the fidelity to the ideal state $\ket{\psi^-}$ of the spin-spin entangled states generated in our setup is $0.92 \pm 0.03$ (\suppl). Combined with measured readout fidelities the generated entangled state is thus expected to violate the CHSH-Bell inequality with $S = 2.30 \pm 0.07$.

\begin{figure}[t]
	\caption{\label{Bell_fig3} \textbf{Characterization of the setup and the entangled state.} \textbf{(a)} Spin-photon correlation measurements. Probability to read out the spin state $\ket{\up}$  at location A (left panel) or B (right panel) when a single photon is detected in the early or late time bin at location C. In the left (right) panel, only emission from A (B) was recorded. Dotted bars show the correlations corrected for finite spin readout fidelity. This yields remaining errors of $1.4 \pm 0.2$\ \% ($1.6 \pm 0.2$\ \%) and $0.8 \pm 0.4$\ \% ($0.7 \pm 0.4$\ \%) when early and late photons are detected at setup A (B). These errors include imperfect rejection of the excitation laser pulses, detector dark counts, microwave pulse errors and off-resonant excitation of the NV. \textbf{(b)} Time-resolved two-photon quantum interference signal. When the NV centres at A and B emit indistinguishable photons (orange), the probability of a coincident detection of two photons, one in each output arm of the beam-splitter at C is expected to vanish.  The observed contrast between the case of indistinguishable versus the case of distinguishable photons of 3 versus 28 events in the central peak yields a visibility of $(90 \pm 6)$\% (\suppl). \textbf{(c)} Characterization of the Bell setup during full operation using (anti-)parallel readout angles. The spins at A (left arrow symbol on the x axis) and B (right arrow symbol) are read out along the $\pm \mathrm{Z}$-axis (left panels), or along the $\pm \mathrm{X}$-axis (right panels). Strong correlations (anti-correlations) are observed for the case where the readout axes are anti-parallel, lower panel (parallel, upper panel), as expected for the $\ket{\psi^-}$ state. The numbers in brackets are the raw number of events. The dotted lines represent the expected correlation based on the measured readout fidelity and the characterisation measurements presented in panels a and b (\suppl).}
\end{figure}

As a final characterization we run the full Bell sequence including random number generation and fast readout, but with co-linear measurement bases ($\mathrm{ZZ}$ and $\mathrm{XX}$) such that spin-spin correlations can be observed with optimal contrast. To test the fast basis selection and rotation the $\mathrm{Z}$ ($\mathrm{X}$) basis measurements are randomly performed along the $+\mathrm{Z}$ ($+\mathrm{X}$) and $-\mathrm{Z}$ ($-\mathrm{X}$) axis. The observed correlations, shown in Fig. 3c (orange bars), are consistent with the estimated quantum state and the independently measured readout fidelities (dotted bars), confirming that the setup is performing as desired.

We find a success probability per entanglement generation attempt of about $6.4 \times 10^{-9}$, yielding a few event-ready signals per hour. Compared to our previous heralded entanglement experiments over 3 meter\ccite{27} this probability is reduced mainly due to additional photon loss (8 dB/km) in the 1.7 km optical fibre. We note that the distance between the entangled electrons is nearly two orders of magnitude larger than in any previous experiment\ccite{30} with entangled matter systems.

Using the results of the characterization measurements we determine the optimal readout bases for our Bell test. A numerical optimization yields the following angles for the readout bases with respect to $\mathrm{Z}$: $0$ (for $a = 0$), $+1/2\pi$ (for $a = 1$), $-3/4 \pi-\epsilon$ (for $b = 0$), $3/4 \pi+\epsilon$ (for $b=1$), with $\epsilon = 0.026\pi$. Adding the small angle $\epsilon$ is beneficial because of the stronger correlations in $\mathrm{ZZ}$ compared to $\mathrm{XX}$. Furthermore, we use the characterization data to determine the time window for valid photon detection events at location C to optimally reject reflected laser light and detector dark counts. We choose this window conservatively to optimize the entangled state fidelity at the cost of a reduced data rate. These settings are then fixed and used throughout the actual Bell test. As a final optimization we replace the photo-detectors at location C with the best set we had available.

We run $245$ trials of the Bell test during a total measurement time of $220$ hours. Figure 4a summarizes the observed data, from which we find $S = 2.42$ in violation of the CHSH-Bell inequality $S \le 2$. We quantify the significance of this violation for two different scenarios (see Figure 4b). First, we analyse the data under the assumptions that the Bell trials are independent of each other, that the recorded random input bits have zero predictability and that the outcomes follow a Gaussian distribution, similar to previous work\ccite{4-9}. This analysis (that we term ``conventional'') yields a standard deviation of $0.20$ on $S$. In this case, the null hypothesis that a Gaussian local realist model for space-like separated sites describes our experiment is rejected with a $p$-value of $0.019$. 

\begin{figure}
	\includegraphics[width=83mm]{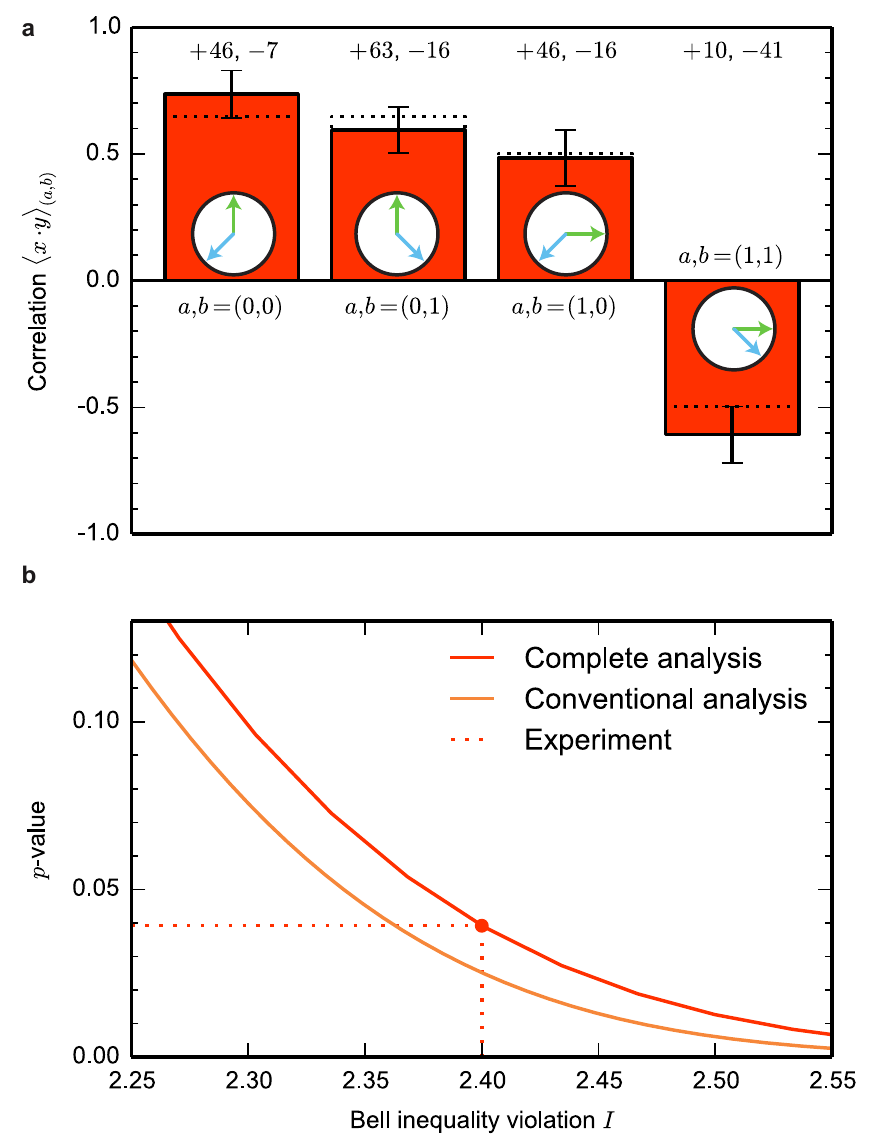}
	\caption{\label{Bell_fig4} \textbf{Loophole-free Bell test results.} \textbf{(a)} Summary of the data and the CHSH correlations. We record a total of $n=245$ trials of the Bell test. The readout bases corresponding to the input values are indicated by the green (for A) and blue (for B) arrows. Dotted lines indicate the expected correlation based on the spin readout fidelities and the characterization measurements presented in Figure 3 (\suppl). Numbers above bars represent the amount of correlated and anti-correlated outcomes respectively. Error bars shown are $\sqrt{1-\left< x \cdot y \right>_{(a,b)}^2/n_{(a,b)}}$, with $n_{(a,b)}$ the number of events with inputs $(a,b)$. \textbf{(b)} Statistical analysis for $n = 245$ trials We test the null-hypothesis that the data can be explained by any local realist model of space-like separated sites, using an analysis allowing arbitrary memory in the devices and taking the measured imperfections of the random number generators into account (Methods). For this analysis, the dependence of the p-value versus the $I$-value is shown (Complete analysis, red). Here $I:= 8 (\frac{k}{n}-\frac{1}{2})$, with $k$ the number of times $(-1)^{(a \cdot b)} x \cdot y=1$. (Note that $I$ equals $S$ from formula \eqref{eq:S_paramter} , for equal $n_{(a,b)}$). A small $p$-value presents strong evidence against the null hypothesis.  In our experiment we find $k = 196$, resulting in a rejection of the null-hypothesis with a $p$-value $\leq 0.039$.  For comparison, we also plot the $p$-value for an analysis (Conventional analysis, orange) assuming independent and identically distributed (i.i.d.) trials, Gaussian statistics, no memory and perfect random number generators. }
\end{figure}

The assumptions made in the conventional analysis are not justified in a typical Bell experiment. For instance, although the locality conditions outlined earlier are designed to ensure independent operation during a \emph{single} trial, the boxes can in principle have access to the entire history including results from all \emph{previous} trials and adjust their output to it\ccite{15,19}. Our second analysis (that we term ``complete'') allows for arbitrary memory, takes the partial predictability of the random input bits into account and also makes no assumption about the probability distributions underlying the data (see Methods). In this case, the null hypothesis that an \emph{arbitrary} local realist model of space-like separated sites governs our experiment is rejected with a $p$-value of $0.039$ (Fig. 4b). 

Our experiment realizes the first Bell test that simultaneously addresses both the detection loophole and the locality loophole. Being free of the experimental loopholes, the setup can test local realist theories of nature without introducing extra assumptions such as fair-sampling, a limit on (sub-)luminal communication or the absence of memory in the setup. Our observation of a loophole-free Bell inequality violation thus rules out \emph{all} local realist theories that accept that the number generators timely produce a free random bit and that the outputs are final once recorded in the electronics. This result places the strongest restrictions on local realistic theories of nature to date. 

Strictly speaking, no Bell experiment can exclude the infinite number of conceivable local realist theories, because it is fundamentally impossible to prove when and where free random input bits and output values came into existence\ccite{12}. Even so, our loophole-free Bell test opens the possibility to progressively bound such less conventional theories: by increasing the distance between A and B (testing e.g. theories with increased speed of physical influence), using different random input bit generators (testing theories with specific free-will agents, e.g. humans), or repositioning the random input bit generators (testing theories where the inputs are already determined earlier, sometimes referred to as ``freedom-of-choice''\ccite{9}). In fact, our experiment already excludes all models that predict that the random inputs are determined a maximum of 690 ns before we record them, because the inequality is still violated for a much shorter spin readout (\suppl).

Combining the presented event-ready scheme with higher entangling rates (e.g. through the use of cavities) provides intriguing prospects for the implementation of device-independent quantum key distribution\ccite{20} and randomness certification\ccite{21,22}. In combination with quantum repeaters, this may enable the realization of large-scale quantum networks that are secured through the very same counter-intuitive concepts that inspired one of the most fundamental scientific debates for 80 years\ccite{1,2}.

\vspace{1cm}
\hrule
\vspace{1cm}

\small
\begin{enumerate}
\item	Einstein, A., Podolsky, B. \& Rosen, N. Can Quantum-Mechanical Description of Physical Reality Be Considered Complete? \emph{Phys. Rev.} \textbf{47}, 777-780 (1935).
\item	Bell, J. S. Speakable and Unspeakable in Quantum Mechanics: Collected Papers on Quantum Philosophy. (Cambridge University Press, 2004).
\item	Freedman, S. J. \& Clauser, J. F. Experimental Test of Local Hidden-Variable Theories. \emph{Phys. Rev. Lett.} \textbf{28}, 938-941 (1972).
\item	Aspect, A., Dalibard, J. \& Roger, G. Experimental Test of Bell's Inequalities Using Time- Varying Analyzers. \emph{Phys. Rev. Lett.} \textbf{49}, 1804-1807 (1982).
\item	Weihs, G., Jennewein, T., Simon, C., Weinfurter, H. \& Zeilinger, A. Violation of Bell's Inequality under Strict Einstein Locality Conditions. \emph{Phys. Rev. Lett.} \textbf{81}, 5039-5043 (1998).
\item	Rowe, M. A. et al. Experimental violation of a Bell's inequality with efficient detection. \emph{Nature} \textbf{409}, 791-794 (2001).
\item	Matsukevich, D. N., Maunz, P., Moehring, D. L., Olmschenk, S. \& Monroe, C. Bell Inequality Violation with Two Remote Atomic Qubits. \emph{Phys. Rev. Lett.} \textbf{100}, 150404 (2008).
\item	Ansmann, M. et al. Violation of Bell's inequality in Josephson phase qubits. \emph{Nature} \textbf{461}, 504-506 (2009).
\item	Scheidl, T. et al. Violation of local realism with freedom of choice. \emph{Proc. Natl. Acad. Sci.} \textbf{107}, 19708-19713 (2010).
\item	Giustina, M. et al. Bell violation using entangled photons without the fair-sampling assumption. \emph{Nature} \textbf{497}, 227-230 (2013).
\item	Christensen, B. G. et al. Detection-Loophole-Free Test of Quantum Nonlocality, and Applications. \emph{Phys. Rev. Lett.} \textbf{111}, 130406 (2013).
\item	Brunner, N., Cavalcanti, D., Pironio, S., Scarani, V. \& Wehner, S. Bell nonlocality. \emph{Rev. Mod. Phys.} \textbf{86}, 419-478 (2014).
\item	Garg, A. \& Mermin, N. D. Detector inefficiencies in the Einstein-Podolsky-Rosen experiment. \emph{Phys. Rev. D} \textbf{35}, 3831-3835 (1987).
\item	Eberhard, P. H. Background level and counter efficiencies required for a loophole-free Einstein-Podolsky-Rosen experiment. \emph{Phys. Rev. A} \textbf{47}, R747-R750 (1993).
\item   Barrett, J., Collins, D., Hardy, L., Kent, A. \& Popescu, S. Quantum nonlocality, Bell inequalities, and the memory loophole. \emph{Phys. Rev. A} \textbf{66}, 042111 (2002).
\item	{\v Z}ukowski, M., Zeilinger, A., Horne, M. A. \& Ekert, A. K. ``Event-ready-detectors'' Bell experiment via entanglement swapping. \emph{Phys. Rev. Lett.} \textbf{71}, 4287-4290 (1993).
\item	Simon, C. \& Irvine, W. T. M. Robust Long-Distance Entanglement and a Loophole-Free Bell Test with Ions and Photons. \emph{Phys. Rev. Lett.} \textbf{91}, 110405 (2003).
\item	Clauser, J. F., Horne, M. A., Shimony, A. \& Holt, R. A. Proposed Experiment to Test Local Hidden-Variable Theories. \emph{Phys. Rev. Lett.} \textbf{23}, 880-884 (1969).
\item	Gill, R. D. Time, Finite Statistics, and Bell's Fifth Position. in Proc. of ``Foundations of Probability and Physics - 2'' \textbf{5}, 179-206 (V{\"a}xj{\"o} Univ. Press, 2003).
\item	Ac{\'i}n, A. et al. Device-Independent Security of Quantum Cryptography against Collective Attacks. \emph{Phys. Rev. Lett.} \textbf{98}, 230501 (2007).
\item	Colbeck, R. Quantum and Relativistic Protocols for Secure Multi-Party Computation. (University of Cambridge, 2007).
\item	Pironio, S. et al. Random numbers certified by Bell's theorem. \emph{Nature} \textbf{464}, 1021-1024 (2010).
\item	Rosenfeld, W. et al. Towards a Loophole-Free Test of Bell's Inequality with Entangled Pairs of Neutral Atoms. \emph{Adv. Sci. Lett.} \textbf{2}, 469-474 (2009).
\item	Gerhardt, I. et al. Experimentally Faking the Violation of Bell's Inequalities. \emph{Phys. Rev. Lett.} \textbf{107}, 170404 (2011).
\item	Robledo, L. et al. High-fidelity projective read-out of a solid-state spin quantum register. \emph{Nature} \textbf{477}, 574-578 (2011).
\item	Barrett, S. D. \& Kok, P. Efficient high-fidelity quantum computation using matter qubits and linear optics. \emph{Phys. Rev. A} \textbf{71}, 060310 (2005).
\item	Bernien, H. et al. Heralded entanglement between solid-state qubits separated by three metres. \emph{Nature} \textbf{497}, 86-90 (2013).
\item	Abell{\'a}n, C., Amaya, W., Mitrani, D., Pruneri, V. \& Mitchell, M. W. Generation of fresh and pure random numbers for loophole-free Bell tests. \emph{ArXiv150602712 Quant-Ph} (2015). at http://arxiv.org/abs/1506.02712
\item	Hong, C. K., Ou, Z. Y. \& Mandel, L. Measurement of subpicosecond time intervals between two photons by interference. \emph{Phys. Rev. Lett.} \textbf{59}, 2044 (1987).
\item	Ritter, S. et al. An elementary quantum network of single atoms in optical cavities. \emph{Nature} \textbf{484}, 195-200 (2012).

\vspace{.5cm}
\hrule
\vspace{.5cm}

\section*{Acknowledgements}

We thank Antonio Ac{\'i}n, Peter Bierhorst, Andrew Doherty, Richard Gill, Peter Gr{\"u}nwald, Marissa Guistina, Laura Mancinska, Hans Mooij, Thomas Vidic, Harald Weinfurter, Yanbao Zhang for fruitful discussions and/or a critical reading of our manuscript and Menno Blauw, Pieter Dorenbos, Rico de Stefano, Christiaan Tiberius, Teus Versluis, Rene Zwagerman and FMVG for help with the realization of the labs and the optical fiber connections. We acknowledge support from the Dutch Organization for Fundamental Research on Matter (FOM), Dutch Technology Foundation (STW), the Netherlands Organization for Scientific Research (NWO) through a VENI grant (THT) and a VIDI grant (SW), the Defense Advanced Research Projects Agency QuASAR program, the Spanish MINECO project MAGO (Ref. FIS2011-23520) and Explora Ciencia (Ref. FIS2014-62181-EXP), the European Regional Development Fund (FEDER) grant TEC2013-46168-R, Fundacio Privada CELLEX and the European Research Council through projects AQUMET and HYSCORE.

\end{enumerate}
\end{document}